# Room temperature ferromagnetism in chemically synthesized ZnO rods

Shalendra Kumar[1*], Y. J. Kim[1], B. H. Koo[1], S. Gautam[2], K. H. Chae[2], Ravi Kumar[3] and C. G. Lee[1*]

[1]School of Nano & Advanced Materials Engineering, Changwon National University, 9 Sarim dong, Changwon 641-773, Korea

[2]Materials Science and Technology Research Division, KIST, Seoul 136-791, Korea

[3]Inter-University Accelerator Centre, Aruna Asaf Ali Marg, New Delhi-110067, India

**Abstract**

We report structural and magnetic properties of pure ZnO rods using X-ray diffraction (XRD), magnetization hysteresis (*M-H*) loop and near edge x-ray fine structure spectroscopy (NEXAFS) study at O K edge. Sample of ZnO was prepared by co-precipitation method. XRD and selective area electron diffraction measurements infer that ZnO rods exhibit a single phase polycrystalline nature with wurtzite lattice. Field emission transmission electron microscopy, field emission scanning electron microscopy micrographs infers that ZnO have rod type microstructures with dimension 200 nm in diameter and 550 nm in length. *M-H* loop studies performed at room temperature display room temperature ferromagnetism in ZnO rods. NEXAFS study reflects absence of the oxygen vacancies in pure ZnO rods.

**\*Corresponding author**

E-mail: **chglee@changwon.ac.kr (C. G. Lee); shailuphy@gmail.com (S. Kumar)**

Ph: +82-55-213-3703; Fax: +82-55-261-7017

## 1. Introduction

In the recent years there has been a significant interest in the development of the room temperature (RT) ferromagnetic semiconductor for spintronics application due to its high Curie temperature above 300K as predicted by theoretical calculations [1]. One of the materials class to realize this function is the dilute magnetic semiconductor (DMSs) where ferromagnetism and semiconductivity gives an additional degree of freedom and functionality for engineering unique device with application ranging from non volatile memory to quantum computing [2-4]. In DMS materials, ZnO based DMSs are particularly promising due to their large band gap, large exciton binding energy (60 meV) over the other conventional wide band gap semiconductors (such as GaN) and lot of potential technological applications in various fields such as: spintronics, light-emitting diode, UV detector, laser diode, cosmetics and biomaterials.

In spite of the progress in the developing ZnO based DMSs, there has been a much controversy concerning the mechanism that causes the magnetism [5-7]. The origin of the ferromagnetism in transition metal doped ZnO is also not yet fully understood and there are debates whether this is caused by defects or doped transition metal ions. Many groups have reported that the ferromagnetism in these materials is due to the defects such as zinc vacancy [8], oxygen vacancy [9], zinc interstitial [10], and oxygen interstitials [11]. Recently some experimental and theoretical groups have reported room temperature ferromagnetism (RTFM) in various un-doped oxides such as $TiO_2$, $HfO_2$, $In_2O_3$ and ZnO. They report that ferromagnetism is also possible in the pure oxides thin film and nano particles. Sundaresan *et al*. [12] reported RTFM in metal oxide nano particle of $CeO_2$, $Al_2O_3$, ZnO, $In_2O_3$, and $SnO_2$, and they explain ferromagnetism in these materials is due to the exchange interactions between localized electron spin moments originating from the surface of nano particles. Recently Wang *et al* [13] reported that Zn vacancy is responsible for the ferromagnetism in ZnO thin films and nano wires. In this letter, we have reported RTFM in pure ZnO rods.

## 2. Experimental

ZnO rods were prepared by the co-precipitation technique using highly pure $Zn(NO_3)_2.6H_2O$ (99.999% from Aldrich) as precursor. Zinc nitrate was dissolved in de-ionized water to a molar concentration of 0.06 M. In this solution, $NH_4OH$ solution was added until the *pH* level reached 9. This mixture was stirred for 3 *h* at RT and then filtered. Precipitate was annealed at 80 $^o$C for 15 *h*. Finally, sample was ground and calcinated at 500 $^o$C for 3 *h*. Philips x-pert x-ray diffractometer with Cu *Kα* ($\lambda$ = 1.54 Å) was used to study single phase nature of the ZnO rods at room temperature. Microstructural analysis of ZnO sample was done using field emission scanning electron microscope (FESEM) and field emission transmission electron microscope (FETEM) (JEM 2100F). Magnetic hysteresis loop measurement was performed at RT using Quantum Design physical properties measurement setup. The near edge X-ray absorption fine structure (NEXAFS) measurement of ZnO at O *K* edge was performed at the soft X-ray beamline 7B1 XAS KIST of the Pohang Accelerator Laboratory (PAL), operating at 2.5 GeV with a maximum storage current of 200 mA. The spectra were collected both in total electron yield (TEY) and fluorescence yield (FY) mode simultaneously at RT in a vacuum better than $1.5 \times 10^{-8}$ Torr. The spectrum in two modes turned out to be nearly identical indicating that the systems are so stable that the surface contamination effects are negligible even in the TEY mode. The spectrum was normalized to incident photon flux and the energy resolution was better than 0.2 eV.

## 3. Results and discussion

Figure 1 shows the X-ray diffraction pattern (XRD) of ZnO rods. The XRD pattern is analyzed and indexed using Powder-X software. XRD pattern indicates that ZnO has single phase nature with wurtzite lattice. The calculated value of the lattice parameters is found 3.254 Å for '*a*' parameter and 5.212 Å for '*c*' parameter.

In order to see the presence of any impurity element in ZnO, we have performed an energy dispersive X-ray (EDX) spectroscopy measurements using FESEM. Figure 2 shows EDX spectrum of ZnO rods. It can be clearly seen from spectrum that the peaks are corresponding to Zn and O which exclude the presence of any impurity element in pure ZnO rods. Inset at top left corner in Figure 3 shows the FESEM micrograph of ZnO rods. FESEM micrograph revealed that ZnO sample has rod type of structure.

Further, in order to see insight of the details of the ZnO rod, we have performed TEM measurements shown as inset in Figure 1. The size of the rod estimated from the TEM micrograph using standard software (*IMAGE J*) is found to be ~200 nm in diameter and ~550 nm in length. One can see from the TEM micrograph that some of the rods are attached sidewise to each other. Inset in Figure 2 illustrate selective area electron diffraction (SAED) obtained by focusing the beam on the ZnO rod. SAED pattern clearly indicate the crystalline nature of each rod. It also shows that rods are indeed in single phase.

Figure 3 shows dc magnetization hysteresis loop of pure ZnO measured at RT. It is observed that ZnO rods exhibit a ferromagnetic behavior at 300K with coercivity of

102 Oe and saturation magnetization of $2.0\times10^{-3}$ emu/g. Some groups have also reported ferromagnetism in pure ZnO nano particles and thin films [12]. They intend that ferromagnetism is the universal feature of the nano particles of non magnetic oxides. They explained that the origin of ferromagnetism in these nanoparticles is due to oxygen vacancies. But recently, some theoretical group has reported that ferromagnetism in ZnO is due to the Zn vacancies and suggested that the origin of the magnetism does not result from the Zn $3d$ electron but it is originated from the unpaired $2p$ electron of O atom in the immediate vicinity of Zn vacancies [13, 14]. Zubiaga *et al* [15] study defect in ZnO bulk using positron annihilation life time spectroscopy and reported the presence of Zn vacancies in pure ZnO. Therefore, the RTFM in pure ZnO rods are due to the Zn vacancies induced spin polarization.

Inset at bottom right corner in Figure 3 shows the O *K*-edge NEXAFS spectrum from pure ZnO. The observed spectral feature can be interpreted as follows: The region between 530 to 538 eV can be ascribed mainly due to O $2p$ hybridization with $4s$ states which form the bottom of the conduction band with peak at ~ 537 eV results due to transition to the non dispersive O $2p_x$ and $2p_{x+y}$ states. The region between 539 to 550 eV can be attributed to O $2p$ hybridized with Zn $4p$ states and above 550 eV the spectrum is due to the hybridization between O $2p$ and Zn $4d$ states [16]. As evident from inset (at bottom right corner) in Figure 3 that the peak (A) at ~532 eV, close to the conduction-band minimum, can be attributed to the hybridization of Zn $3d$ states with O $2p$ states. In addition, there is no spectral feature between 535 to 540 eV which results due ***to*** the presence of oxygen vacancies. Therefore, absence of spectral features at ~537 eV indicates the oxygen vacancies related defects are not present in the system. Therefore, absence of O vacancies in pure ZnO rods infers that RTFM in ZnO may be due to Zn vacancies.

## 4. Conclusions

In summary, we have successfully synthesis single phase ZnO rods. XRD and SAED study shows single phase nature of ZnO rods. NEXAFS at O *K* indicate the absence of oxygen vacancies. Magnetic study infers RTFM in ZnO rods.

## Acknowledgements

This work was supported by the Korean Research Foundation Grant funded by the Korean Government (MOEHRD) (KRF-2008-005-J02703).

## References

[1] Dietl T, Ohno H, Matsukura F, Cibert J, Ferrand D, Science 2000; 287: 1019-22

[2] Fukumura T, Jin Z, Ohtomo A, Koinuma H, and Kawasaki M, Appl Phys Lett 1999; 75: 3366-68

[3] Jung SW, An SJ, Yi GC, Jung CU, Lee SI, and Cho S, Appl Phys Lett 2002; 80: 4561-63

[4] Ohno H, Science 1998; 281: 951-56

[5] Ueda K, Tabata H, and Kawai T, Appl Phys Lett 2001; 79: 988-90

[6] Rode K, Anane A, Mattana R, Contour JP, Durand O, LeBourgeois R, J Appl Phys 2003; 93: 7676-78

[7] Venkatesan M, Fitzgerald CB, Lunney JG, Coey JMD, Phys Rev Lett 2004; 93:177206-4

[8] Tuomisto F, Ranki V, Saarinen K, Look DC, Phy Rev Lett 2003; 91: 205502-4

[9] Mahan GD, J Appl Phys 1983; 54: 3825-32

[10] Neumann G, Phys Status Solidi B 1981; 105: 605-12

[11] Brauer G, Anwand W, Skorupa W, Kuriplach J, Melikhova O, Moisson C, Wenckstern HV, Schmidt H, Lorenz M, Grundmann M, Phy Rev B 2006; 74: 045208-10

[12] Sundaresan A, Bhargavi R, Rangarajan N, Siddesh U, Rao CNR, Phy Rev B 2006; 74: 161306(R)-4

[13] Wang Q, Sun Q, Chen G, Kawazoe Y, Jena P, Phy Rev B 2008; 77: 205411-7

[14] Hong NH, Sakai J, Brize V, J Phys: Condens. Matter, 2007; 19: 036219-6

[15] Zubiaga A, Plazaola F, Garcia JA, Tuomisto F, Muñoz-Sanjosé V, Tena-Zaera R, Phys. Rev. B, 2007; 76: 085202-8

[16] Guo JH, Vayssieres L, Persson C, Ahuja R, Johnsson B, Nordgren J, J Phys: Condens. Matter 2002; 14: 6969-6

## Figure Captions

**Fig.1.** X-ray diffraction pattern of ZnO. Inset shows TEM micrograph of ZnO rods.

**Fig.2.** EDAX spectrum obtained from of ZnO rods. Inset shows SAED pattern of ZnO rods.

**Fig.3.** Room temperature hysteresis curve obtained from ZnO rods. Inset at top left corner shows FESEM micrograph and inset at bottom right corner shows NEXAFS spectrum at O $K$-edge of ZnO rods

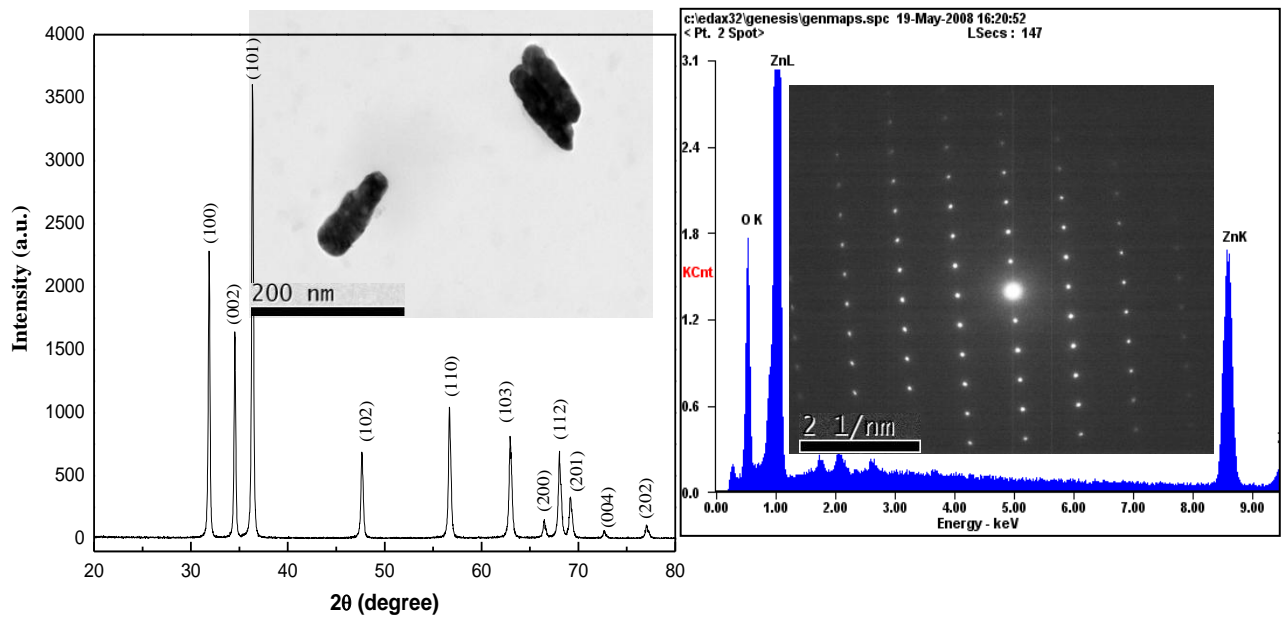

Fig. 1 Fig. 2

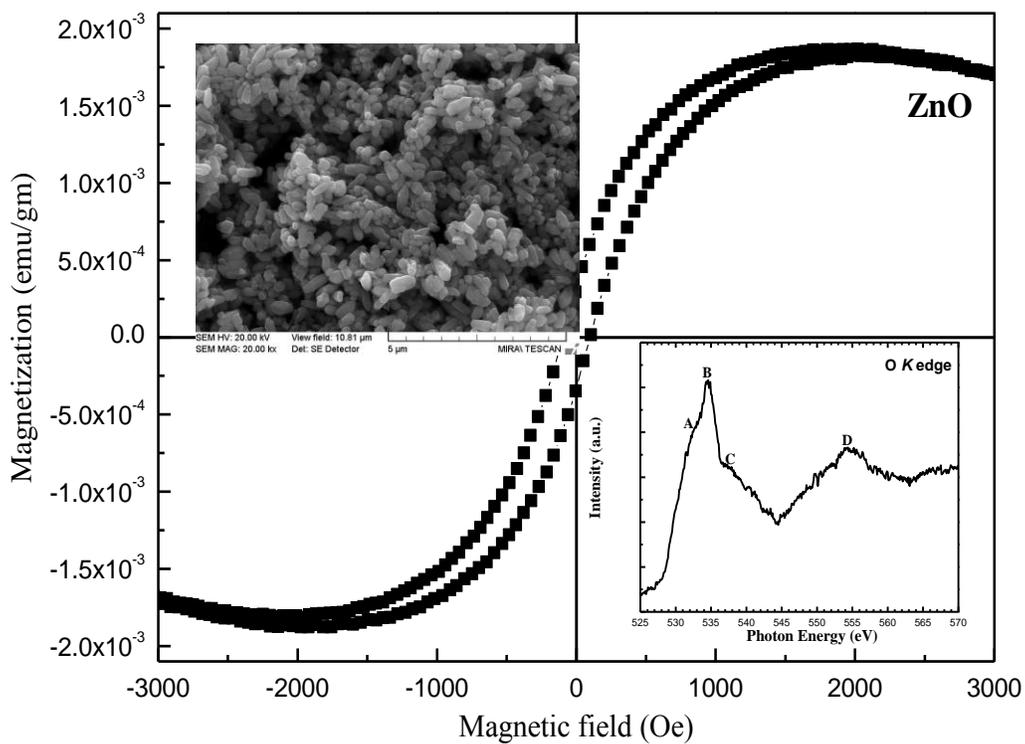

Fig. 3